\documentclass[amsmath,amssymb,pra,twocolumn,showpacs]{revtex4}
\usepackage{graphicx}
\usepackage{dcolumn}
\usepackage{bm}
\usepackage{epsfig}
\usepackage{bm}

\newcommand{\bear}{\begin{eqnarray}}
\newcommand{\enar}{\end{eqnarray}}
\begin{document}

\title{Polynomial method to study the entanglement of pure $N$-qubit states}

\author{H. M\"akel\"a and A. Messina}

\affiliation{Dipartimento di Scienze Fisiche ed Astronomiche, Universit\`a di Palermo, via Archirafi 36, I-90123 Palermo, Italy}

\begin{abstract}
We present a mapping which associates pure $N$-qubit states with a polynomial. 
The roots of the polynomial characterize the state completely. Using the properties of the polynomial  
we construct a way to determine the separability and the number of unentangled qubits of pure $N$-qubit states.
\end{abstract}
\maketitle

\maketitle                             
\section{Introduction}
Considerable effort is spent in developing methods for the detection 
and classification of entangled states. One important aim is to find 
ways to detect the separability of mixed states consisting of an arbitrary 
number of subsystems. While a general, easily computable, method to detect 
the separability of arbitrary mixed multipartite states is still 
lacking, some partial results exist. Maybe the most famous separability 
condition for mixed states is the positive partial transposition, also known as  
Peres-Horodecki criterion \cite{Peres96,Horodecki96}. 
This method is simple and easy to apply, 
but it can be used  to detect only bipartite separability. 
Therefore various separability conditions which work in an $N$-party 
setting have been developed. Examples of these are permutation criteria, 
where the indices of the density matrix are permuted  \cite{Horodecki06}, 
the use of quadratic Bell-type inequalities \cite{Seevinck08}, 
algorithmic approaches \cite{Doherty05}, and the use of positive maps \cite{Horodecki01}.  
For a more comprehensive list, see \cite{Guhne09,Horodecki09}. 
In the case of pure states the situation is simpler.  
A pure $N$-partite state is separable if  and only if all the reduced 
density matrices of the elementary subsystems describe pure states.  
Alternatively, in a bipartite case, separability can  be determined by 
calculating the Schmidt decomposition of the state.   
Unfortunately, the concept of the Schmidt decomposition cannot be 
straightforwardly generalized to the case of $N$ separate subsystems \cite{Peres95,Thapliyal99}.  
In addition to these two well-known methods, various other approaches to 
the pure state separability have been discussed. 
A separability condition based on comparing the amplitudes and phases of the 
components of the state has been discussed in \cite{Jorrand03,Matsueda07}. 
It has been shown that the separability of pure three-qubit states 
can be detected by studying two-qubit density operators \cite{Brassard01} 
and expectation values of spin operators \cite{Yu05,Yu07a}. 
Separability tests based on studying matrices constructed from 
the components of the state vector, known as coefficient matrices, 
have gained attention recently \cite{Lamata06,Li08,Huang09}.

In this article we present a mapping which associates the pure states of an $N$-qubit system 
with a  polynomial. The roots of the polynomial determine the state completely and vice versa. 
We show that this polynomial establishes a simple way to test the separability of pure $N$-qubit states 
and to study the number of unentangled particles. 
The idea to associate a state of a quantum mechanical system with a polynomial is not new. 
Already in 1932 E. Majorana presented a polynomial, nowadays known as the Majorana polynomial, 
which he used to show that the states of a spin-$S$ particle can 
be expressed as a superposition of symmetrized states of $2S$ spin-$\frac{1}{2}$ systems \cite{Majorana32,Bloch45}.   
This decomposition, the Majorana representation, has been relatively unknown for a long time.  
However, it has recently found applications in may different fields, such as in studying the symmetries of 
spinor Bose-Einstein condensates   \cite{Barnett06,Barnett07,Makela07,Barnett09},  
in the context of reference frame alignment \cite{Kolenderski08}, 
in helping to define anticoherent spin states \cite{Zimba06}, and in calculating the 
spectrum of the Lipkin-Meshkov-Glick model \cite{Ribeiro07,Ribeiro08}. 
It has also been used to give a graphical representation for the states of an $n$-level system \cite{Bijurkar06}. 

The states of an $N$-qubit quantum register can 
be viewed as the spin states of a particle with 
spin $S=(2^N-1)/2$. Therefore, expressing the pure 
states of an $N$-qubit system utilizing the  
Majorana representation requires the use of $2^N-1$ spin-$\frac{1}{2}$ systems. 
In the approach we present in this article only $N$ two-level systems 
are needed to characterize the states of this system. 
The Majorana representation is useful in studying the 
behavior of spin states under spin rotations as a spin rotation 
of a spin-$S$ particle is equivalent with rotating the states 
of the constituent spin-$\frac{1}{2}$ particles \cite{Bloch45}. 
However, when discussing the states of an $N$-qubit 
quantum register, this property is not very helpful and 
therefore the benefits of the Majorana representation cannot 
fully be taken advantage of. In this case the simplified 
description presented in this article becomes useful.

This article is organized as follows. 
In Sec. II we introduce a mapping between the pure states of an $N$-qubit 
quantum register and polynomials. We argue that the roots of a polynomial 
determine a unique state and vice versa.      
In Sec. III we calculate the polynomial of separable  pure states and derive a necessary and sufficient 
condition for the separability of an arbitrary pure $N$-qubit state.  
We also briefly discuss the generalization of the polynomial approach 
to systems containing $N$ copies of an $h$-level system. 
In Sec. IV  we show how the polynomial can be used to study the number of unentangled qubits. 
In Sec. V we present the conclusions.   

\section{Characteristic polynomial}
We denote the basis of the qubit $j$ by $\{|0\rangle_j,|1\rangle_j\}$, so the basis vectors 
of an $N$-qubit quantum register can be chosen as $|i_0i_1\cdots i_{N-1}\rangle\equiv|i_0\rangle_0\otimes|i_1\rangle_1\otimes\cdots\otimes |i_{N-1}\rangle_{N-1}$, where every $i_j\in \{0,1\}$.    
Each natural number $0\leq i\leq 2^N-1$ can be written using binary notation as 
$i=\sum_{j=0}^{N-1} i_j 2^j$, where $i_j\in \{0,1\}$. Using this we can
associate the basis vector $|i_0i_1\cdots i_{N-1}\rangle$  
 with $|i\rangle_d$. Here the subscript $d$ shows 
that decimal notation is used to label the basis states. Let 
\begin{equation}\label{somestate}
\phi=\sum_{i=0}^{2^N-1} \, C_i|i\rangle_d, \quad C_i\in \mathbb{C},  
\end{equation}
be some, possibly unnormalized, state vector of an $N$-qubit system.  
We associate this vector with the polynomial
\begin{equation}\label{poly}
P(\phi;x)\equiv \sum_{i=0}^{2^N-1} C_i x^i,
\end{equation}
which we call the characteristic polynomial of $\phi$. 
By the fundamental theorem of algebra, this polynomial can be written in a unique way as 
\begin{equation}
\label{P}
P(\phi;x)=C_k \prod_{j=0}^{k-1}(x-x_j),
\end{equation}
where $\{x_j\,|\,j=0,1,\ldots,k-1\}$ are the roots and $k$ is the degree of $P(\phi;x)$. 
If $k=0$ we define $\prod_{j=0}^{-1}(x-x_j)=1$. 
The set of vectors  $\{c\,\phi \,|\, c\in\mathbb{C}, c\not=0\}$ 
determines a unique set of roots and each set of roots 
$\{x_0,x_1,\ldots ,x_{k-1}\}$ determines the vector $\phi$ up to 
normalization and phase. Therefore we have a bijective map between 
the pure states of an $N$-qubit quantum register and the roots of 
complex polynomials of degree $k\leq 2^{N}-1$ 
\footnote{We could naturally define the characteristic polynomial as 
$P(\phi;x)\equiv \sum_{i=0}^{2^N-1} C_i g_i x^i$, 
where $\{g_i\}$ is a set of $2^N$ arbitrarily chosen nonzero complex numbers. However, 
in order to simplify the ensuing calculations we 
choose $g_i=1$ for each $i$. The choice $g_i={2^N-1\choose i}^{1/2}$ 
corresponds to the Majorana representation, 
see Refs. \cite{Majorana32,Bloch45}.  \label{footnote}}. Explicitly, the 
components of $\phi$ are determined by the roots through the formula
\begin{equation}\label{components}
C_i=(-1)^{k-i}\sum_{j_0<j_1<j_2<\cdots <j_{k-1-i}}
\!\!\!\!\!\!\!\!\!\!\!\!x_{j_0}x_{j_1}x_{j_2}\cdots x_{j_{k-1-i}},
\end{equation}
where $i=0,1,2,\ldots,k-1$ and we have chosen $C_k=1$. 
The roots contain the same amount of information on the system 
as the state vector $\phi$. In particular, all the entanglement 
properties of $\phi$ are encoded in the set of roots corresponding to $\phi$.  
With the help of the roots the state $\phi$ can be given a geometrical representation as  
$2^N-1$ points on the Bloch sphere, see Ref. \cite{Makela09}.

\section{Separable pure $N$-qubit states}
In this section we show how the separability of $\phi$ can be detected 
with the help of $P(\phi;x)$. In order to do so, we first calculate 
the characteristic polynomial of a separable state. 
Any separable pure state $\phi_{\textrm{s}}$ can be written as 
\begin{align}
\label{Productstate}
\nonumber
\phi_{\textrm{s}}&=\bigotimes_{j=0}^{N-1}\phi_j\\
&=\bigotimes_{j=0}^{N-1}(a_j |0\rangle_j +b_j |1\rangle_j)\quad a_j,b_j\in\mathbb{C}.
\end{align}
Assume that $|l\rangle_d$ is a basis state of an $L$-qubit system 
and that $|m\rangle_d$ is that of an independent $M$-qubit system. 
Using the binary expressions for $l$ and $m$ it is easy to see that  
\begin{equation}\label{tworegs}
|l\rangle_d |m\rangle_d =|l+2^L m\rangle_d 
\end{equation}
holds for the tensor product of $|l\rangle_d$ and $|m\rangle_d$.  
Here and in what follows we omit the tensor product symbol. 
Let  $\xi^L$ and $\xi^M$ be states of $L$-qubit and $M$-qubit 
quantum registers, respectively. 
Then we can write $\xi^L=\sum_{i=0}^{2^L-1}\xi^L_i |i\rangle_d$ 
and $\xi^M=\sum_{i'=0}^{2^M-1}\xi_{i'} |i'\rangle_d$. 
If $\phi\in  (\mathbb{C}^2)^{L+M}$ can be written as $\phi=\xi^L\xi^M$, then
\begin{align}
\phi &=\sum_{i=0}^{2^L-1}\sum_{i'=0}^{2^M-1}\xi_i^L  \xi_{i'}^M|i\rangle_d |i'\rangle_d\\
&= \sum_{l=0}^{2^L-1}\sum_{i'=0}^{2^M-1}\xi_i^L  \xi_{i'}^M|i+2^L i'\rangle_d, 
\end{align}
where we have used Eq. (\ref{tworegs}). Consequently, the characteristic polynomial of $\phi$ becomes  
\begin{align}
\nonumber
P(\phi;x)&=\sum_{i=0}^{2^L-1}\sum_{i'=0}^{2^M-1}\xi^L_i  \xi_{i'}^M x^{i+2^L i'}\\
\nonumber
&=\Big(\sum_{i=0}^{2^L-1}\xi_i^L x^i\Big) \sum_{i'=0}^{2^M-1}\xi_{i'}^M (x^{2^L})^{i'}\\
&= P(\xi^L;x)P(\xi^M;x^{2^L}).\label{ProdPoly}
\end{align}  
Therefore, if the state of the quantum register is the product 
of an $L$-qubit state and an $M$-qubit state, the characteristic 
polynomial factorizes as the product of the polynomials of the two states. 
In the polynomial of the $M$-qubit state the variable $x$ is replaced by $x^{2^L}$. 
Using Eq. (\ref{ProdPoly}) it is easy to calculate the characteristic 
polynomial $P(\phi_{\textrm{s}};x)$ of a separable state 
$\phi_{\textrm{s}}\equiv\phi_0\phi_1\cdots\phi_{N-1}$ given by Eq. (\ref{Productstate}). 
By defining $\phi_{j;N}\equiv \phi_j\phi_{j+1}\cdots\phi_{N-1}$, 
so that $\phi_{j;N}=\phi_j\phi_{j+1;N}$, and  
using Eq. (\ref{ProdPoly}) repeatedly  we get 
\begin{align}
\nonumber
\label{Poly}
P(\phi_{\textrm{s}};x)
&=P(\phi_0,x)P(\phi_{1;N},x^2) \\
\nonumber
&=P(\phi_0,x)P(\phi_{1},x^2 )P(\phi_{2;N},x^4 )\\
\nonumber
&=P(\phi_0,x)P(\phi_{1},x^2)P(\phi_{2},x^4)P(\phi_{3;N},x^8)\\
\nonumber
&=\cdots \\
\nonumber
&=\prod_{j=0}^{N-1} P(\phi_j,x^{2^{j}})\\
&=\prod_{j=0}^{N-1} (a_j+b_j x^{2^{j}}).
\end{align}
We see that the characteristic polynomial of a separable state can always be written in the form of (\ref{Poly}). 
On the other hand, there always exists a separable   
state whose characteristic polynomial is given by Eq. (\ref{Poly}), namely the state $\phi_{\textrm{s}}$. 
From the definition of $P(\phi;x)$ it follows that 
if $P(\phi;x)=P(\tilde{\phi};x)$, then necessarily $\phi=\tilde{\phi}$. Therefore $\phi_{\textrm{s}}$ 
is the unique vector which gives rise to the polynomial of Eq. (\ref{Poly}). 
In conclusion, 
a pure $N$-qubit state $\phi$ is separable if and only if $P(\phi;x)$ can be written as in Eq. (\ref{Poly}). 
The roots of this equation are 
\begin{equation}
\label{roots}
x_{jm}=\left(-\frac{a_j}{b_j}\right)^{1/2^j}e^{i\frac{2\pi m}{2^j}},\quad m=0,1,\ldots ,2^j-1,
\end{equation}
where $b_j$ has to be nonzero. If $b_j$ is zero the degree of the polynomial is decreased 
by $2^j$ from the maximal degree $2^N-1$.

The separability of a state $\phi$ can be determined by   
calculating the roots of $P(\phi,x)$ and checking if they are of the form given by 
Eq. (\ref{roots}). These calculations can in practice turn out to be very complicated. 
It may be computationally demanding to achieve accurate enough results in order to reliably 
see how the roots are distributed in the complex plane.  
This is partly related to the fact the degree of the polynomial $P(\phi;x)$ can be $2^N-1$, 
which grows rapidly with $N$, rendering the calculation of roots time-consuming for large $N$.  
However, we will show next that the roots of $P(\phi;x)$ can be expressed in a 
simple way in terms of the components $\{C_i\}$ of the state vector if $\phi$ is separable.
Let $\phi_{\textrm{s}}$ be the separable state given by Eq. (\ref{Productstate}).
When this vector is written in the form $\phi_{\textrm{s}}=\sum_{i=0}^{2^N-1} \, C_i|i\rangle_d$,  
the components $C_i$  are easily obtained by noting that $i_j=0$ ($i_j=1$) corresponds to $a_j$ ($b_j$): 
\begin{equation}\label{ck}
C_i=\prod_{j=0}^{N-1}[(1-i_j)a_j+i_j b_j]. 
\end{equation}
Here we have used the binary form of $i$, that is, we have written $i=\sum_{j=0}^{N-1}i_j 2^j$. 
We assume that $C_k\not=0,C_{k+1}=\cdots=C_{2^N-1}=0$, 
so that the degree of $P(\phi_{\textrm{s}};x)$ is $k$. 
By writing $k=\sum_{j=0}^{N-1} k_j 2^j$ we see that if $k_j=1$, 
then $(k-2^{j})_{l}=k_l-\delta_{jl}$, $l=0,1,\ldots ,N-1$. Using this and Eq. (\ref{ck}) it is easy to see that now  
$a_j/b_j=C_{k-2^j}/C_k$. On the other hand, if $k_j=0$, then $(k+2^j)_l=k_l+\delta_{jl}$ and Eq. (\ref{ck}) gives  
$b_j/a_j=C_{k+2^j}/C_k=0$. 
Summarizing, 
\begin{equation}
\label{ratios}
\begin{array}{llll}
\dfrac{a_j}{b_j}&=&\dfrac{C_{k-2^j}}{C_k}  &\qquad\textrm{if } k_j=1,\\ 
b_j&=& 0 &\qquad\textrm{if } k_j=0.
\end{array}
\end{equation}
Using Eq. (\ref{roots}) we immediately see that the $k$ roots of $P(\phi_{\textrm{s}};x)$ are 
\begin{equation}
\label{xlm}
x_{jm}=\left(-\dfrac{C_{k-2^j}}{C_k}\right)^{1/2^j}e^{i\frac{2\pi m}{2^j}},\quad m=0,1,\ldots ,2^j-1,
\end{equation}
where $j$ takes those values for which $k_j =1$.  
On the other hand, if the roots and their multiplicities are known, the polynomial 
can be determined up to a multiplying constant. In particular, if $x=0$ is a root, then 
its multiplicity has to be equal to the lowest power of the polynomial. 
In conclusion, an arbitrary pure state $\phi$ is separable  
if and only if 
\begin{equation}
\begin{array}{lll}
&\textrm{(Ia) }& \textrm{All the numbers } x_{jm} \textrm{ given by Eq. (\ref{xlm}) are}\\&&\textrm{roots of } P(\phi;x).\\ 
&\textrm{(Ib) }& \textrm{The number of } x_{jm} \textrm{ equaling zero is equal to}\\&&\textrm{the lowest power of } P(\phi;x). 
\nonumber 
\end{array}
\end{equation}
An alternative formulation is that $\phi$ is separable if and only if the quantity 
\begin{equation}
S(\phi)\equiv\sum_{j,\, k_j=1}\sum_{m=0}^{2^j-1}|P(\phi;x_{jm})|
\end{equation} 
equals zero and Condition (Ib) holds. Note that 
if $k=0$ the state is separable.

If a state is found to be separable, then the one-particle states it 
consists of can be explicitly constructed with the help of the ratios $a_j/b_j$ given by Eq. (\ref{ratios}). 
We now present some examples of the detection of separability of states with several freely varying components.

\subsection{Example 1} 
As the first example we consider a state defined as
\begin{equation}
\xi^N=C_0|0\rangle_d+C_1|1\rangle_d+\cdots +C_{k-2}|k-2\rangle_d+C_k|k\rangle_d,
\end{equation}
where $C_0,C_k\not=0$ and $k$ is odd.  
Since $k$ is odd $k_0=1$ and Eq. (\ref{xlm}) shows that $x_{00}=C_{k-1}/C_k=0$. Because 
$P(\xi^N;x_{00})=C_0\not=0$, $\xi$ cannot be a separable state. 
In a three-qubit case we see that, for example, 
\begin{align}
\nonumber 
\xi^3 &= C_0|000\rangle+C_1|100\rangle+C_2|010\rangle+C_3|110\rangle\\
&+C_4|001\rangle+C_5|101\rangle+C_7|111\rangle
\end{align}
where $C_0 C_7\not =0$ cannot be separable.

In order to compare our approach with other separability tests, we now check the separability 
of $\xi^3$ using an alternative method. There exist various (partial) multipartite 
separability criteria for mixed states (see, for example, \cite{Horodecki06,Seevinck08,Doherty05,Horodecki01}). 
While these are useful when mixed states are studied, in the case of pure states the most convenient 
separability check is usually the standard method of calculating the reduced single-qubit 
density matrices of the $N$-qubit state. This view is supported by the fact that 
alternative pure state separability tests require examining the properties of matrices 
that are higher dimensional than the  two-by-two dimensional reduced 
single-qubit density matrices \cite{Brassard01,Huang09} or   
require the calculation of the expectation values of operators expressed as tensor 
products of the Pauli spin matrices \cite{Yu05,Yu07a}. 
This results in a complex calculation if a state that contains many freely  
varying components, such as $\xi^3$, is studied.  
For these reasons we now examine the separability of $\xi^3$ by using the method of partial traces. 
Here and in what follows we denote the reduced single-qubit density matrix pertaining to qubit $j$ 
by $\rho_j$. Now the indexing of qubits runs from $0$ to $N-1$. 
The vector $\xi^3$ is separable if and only if any two of the three 
 density matrices $\rho_0,\rho_1$, and $\rho_2$ describe pure states. 
The state $\rho_j$ is pure if and only if $\det(\rho_j)=0$, so 
if $\det(\rho_j)\not=0$ for at least one $j$, then $\xi^3$ is entangled. 
As an example we determine $\det(\rho_0)$. A simple calculation shows that  
the single-qubit reduced density matrix of the first qubit is
\begin{equation} 
\footnotesize
\rho_0=\left(\begin{array}{cc}
|C_0|^2+|C_2|^2+|C_4|^2 & C_0 C_1^*+C_2 C_3^*+C_4C_5^*\\
 C_0^* C_1+C_2^* C_3+C_4^*C_5 & |C_1|^2+|C_3|^2+|C_5|^4+|C_7|^2
\end{array}\right)
\end{equation}
Using the inequality $\textrm{Re}(C)\leq |C|$, where $C$ is an arbitrary 
complex number, it can be shown that the following inequality holds 
for the determinant of $\rho_0$
\begin{align}
\footnotesize
\det(\rho_0)
\nonumber 
&\geq |C_7|^2(|C_0|^2+|C_2|^2+|C_4|^2)+ (|C_0C_3|-|C_1C_2|)^2\\
&+ (|C_0C_5|-|C_1C_4|)^2+(|C_2C_5|-|C_3C_4|)^2. 
\end{align}
This is bounded below by $|C_0C_7|^2>0$, confirming the aforementioned result concerning the 
separability of $\xi^3$. 
Therefore a necessary condition for the separability of $\xi^3$ 
can be  straightforwardly obtained using partial traces.  
However, the polynomial method provides a simpler separability test in the present example.    
Even more so if instead of $\xi^3$ the separability of the 
$N$-qubit state $\xi^N$ is studied.

\subsection{Example 2}
In the second example we choose $\xi^N$ such that the degree of $P(\xi^N;x)$ is $k=2^N-2$. Then  
$k_j=1-\delta_{0j}$. We assume that $C_{2^{N-1}-2}(=C_{k-2^{N-1}})=0$, from which it follows that 
 $x_{(N-1)m}=0$ for $m=0,1,\ldots ,2^{N-1}-1$. 
According to Condition (Ib) the lowest order of the polynomial has to be at least $2^{N-1}$ for the state to be a product state.  
Thus, if $C_i\not=0$ for at least one $i$ such that $0\leq i <2^{N-1}$,  $i\not= 2^{N-1}-2$, then $\xi^N$ must be entangled. 
In the case of a three-qubit system this result means that 
\begin{eqnarray}
\nonumber
\xi^3 &=&C_0|000\rangle+C_1|100\rangle+C_3|110\rangle+C_4|001\rangle\\
&&+C_5|101\rangle+C_6|011\rangle, \quad C_6\not =0,
\end{eqnarray}
cannot be a product state if $C_0,C_1,$ or $C_3$ is nonzero.  
If $C_4=0$ we have $x_{10}=x_{11}=0$, which means that all $x_{jm}$ are equal to zero. Then 
$\xi^3$ cannot be separable unless all $C_i$ except $C_6$ are zero. 
The reduced single-qubit density matrices $\rho_0,\rho_1$, and $\rho_2$ can be straightforwardly calculated and 
are not presented here. The determinant of $\rho_2$ is
\begin{align}
\nonumber
\det(\rho_2)&=(|C_0|^2+|C_1|^2+|C_3|^2)(|C_4|^2+|C_5|^2+|C_6|^2)\\
&-|C_0 C_4^*+C_1C_5^*|^2\\
\nonumber 
&\geq  |C_6|^2(|C_0|^2+|C_1|^2+|C_3|^2)
\\
\label{ex2}
&+|C_3|^2(|C_4|^2+|C_5|^2) +(|C_0 C_5|-|C_1 C_4|)^2, 
\end{align} 
where we have obtained a lower bound for the determinant in the same fashion 
as in the previous example.  
We reproduce the earlier result that $\xi^3$ is necessarily entangled if $C_1,C_2$ or $C_3$ is nonzero. 
In order to determine the separability conditions in the case $C_4=0$ one
has to calculate $\det(\rho_0)$ and repeat the above calculation for this quantity. The result agrees with the 
one obtained using the polynomial approach, that is,  if $\xi^3$ is separable and $C_4=0$, 
 then only $C_6$ can be nonzero. We see that also in this case the polynomial approach  
provides an easier way to check the separability than the method of partial traces. 

\subsection{Example 3}
As the final example we study a state given by
\begin{align}
\nonumber\label{xi}
\xi^N &=\sum_{\substack{i=1\\i\not =0,4,8,\ldots,2^N-4}}^{2^N-1}|i\rangle_d+e^{i\theta}\sum_{i=0}^{2^{N-2}-1}|4i\rangle_d\\
&=\sum_{i=0}^{2^N-1}|i\rangle_d+\left(e^{i\theta}-1\right)\sum_{i=0}^{2^{N-2}-1}|4i\rangle_d. 
\end{align}
Now $C_{(2^N-1)-2^j}/C_{2^N-1}=1$ for all $j$, so Eq. (\ref{roots}) gives  
\begin{equation}
x_{jm}=e^{i\frac{(2m+1)\pi}{2^j}},\quad m=0,1,\ldots ,2^j-1,
\end{equation}
where $j=0,1,2,\ldots, N-1$. 
Using the sum formula of geometric series we find that the characteristic polynomial can 
be written as
\begin{equation}
\label{Pxi}
P(\xi^N;x)=\frac{x^{2^N}-1}{x-1}+(e^{i\theta}-1)\frac{x^{2^N}-1}{x^4-1}. 
\end{equation}
It is easy to see that for $j=2,3,\ldots,N-1$ 
\begin{equation}
\label{j2}
P(\xi^N;x_{jm})=0,\quad m=0,1,\ldots,2^{j}-1,
\end{equation}
while
\begin{equation}\label{j0}
P(\xi^N;x_{00})=P(\xi^N;x_{10})=P(\xi^N;x_{11})=2^{N-2}(e^{i\theta}-1). 
\end{equation}
The state $\xi^N$ is separable if and only if $\theta=2\pi n$ for some integer $n$. 
If $\xi^N$ is separable Eq. (\ref{ratios}) shows that  
$\xi^N=\otimes_{j=0}^{N-1}(|0\rangle_j+|1\rangle_j)$.

Now the $N$ reduced single-qubit density matrices of $\xi^N$ 
can be straightforwardly determined. Lengthy calculation shows 
that $\textrm{det}(\rho_0)=\textrm{det}(\rho_1)=2^{2N-3}(1-\cos\theta)$ 
and $\textrm{det}(\rho_j)=0$ when $j=2,3,\ldots, N-1$, confirming the earlier result.  
In the present example the polynomial method does not seem to provide  
as obvious calculational simplification as in the previous two examples. 

\subsection{Generalization to $h$-level systems}
We now briefly discuss a generalization of the separability test 
 to a system consisting of $N$ copies of an $h$-level system. 
We write the basis of a single $h$-level system as 
$\{|0\rangle_h,|1\rangle_h,\ldots,|h-1\rangle_h\}$ 
and choose the basis vectors for the $N$-partite system as   
$|i\rangle_d=|i_0i_1\cdots i_{N-1}\rangle_h$, where  
$i=\sum_{j=0}^{N-1}i_j h^j$ and $i_j\in \{0,1,2,\ldots, h-1\}$.  
An arbitrary pure state can be expressed as 
\begin{eqnarray}
\label{initialbasis}
\phi =\sum_{i=0}^{h^N-1} C_i|i\rangle_d. 
\end{eqnarray}
Let $\phi_{\textrm{s}}^h=\phi_0^h\phi_1^h\cdots\phi_{N-1}^h$ be a separable state   
where $\phi_j^h=a_j|0\rangle_h+b_j|1\rangle_h+c_j|2\rangle_h+\cdots +q_j|h-1\rangle_h$. 
A straightforward calculation shows that 
\begin{eqnarray}
\label{Ph}
&&P(\phi_{\textrm{s}}^h;x)=
\!\!\prod_{j=0}^{N-1} \left(a_j+b_j x^{h^{j}}+\cdots +q_j x^{(h-1)h^{j}}   \right).
\end{eqnarray}
In order to establish a separability test, one has to express the roots of this 
polynomial in terms of the coefficients $C_0,C_1,\ldots,C_{h^N-1}$. 
This is possible but complicated if $2<h<6$. If $h\geq 6$, the roots 
cannot be calculated analytically and therefore cannot be
written using the coefficients $C_i$. Thus the separability test 
can be extended to systems containing less than six levels, but it is  
more complicated to apply than in the two-level case.  
An extension is not feasible if the number 
of levels is equal to or larger than six.

\section{Number of unentangled qubits}

Entangled states can be classified based on the number of unentangled one-qubit states. 
The state $\phi$ is said to contain $n$ unentangled qubits if it can be written 
as a product of $n$ single-qubit states $\phi_l $ and an $(N-n)$ -qubit state $\phi^{N-n}$. 
In order to study the number of unentangled particles, 
we determine the characteristic polynomial of a state which separates as a product 
of a one-qubit state and an $(N-1)$-qubit state.  
We write $\phi=\phi_j \phi^{N-1}$, where 
$\phi_j=a_j|0\rangle_j+b_j|1\rangle_j$ is the state of the qubit $j$ 
and $\phi^{N-1}$ gives the state of the  rest of the qubits. 
As before, the degree of the polynomial is denoted by $k$.  
Using Eq. (\ref{tworegs}) we see that the characteristic polynomial 
of the basis states reads 
\begin{align}
\nonumber 
P(|i\rangle_d;x)&=P(|i_0 i_1\cdots i_{N-1}\rangle;x)\\
\label{Pk}
&= x^{i_0 2^0}x^{i_1 2^{1}} x^{i_2 2^{2}} \cdots x^{i_{N-1} 2^{N-1}}. 
\end{align} 
We write the $(N-1)$-qubit state as 
\begin{equation}
\phi^{N-1}=\!\!\!\!\!\!\!\!\!\!\sum_{i_l \in\{0,1\},l\not =j }
\!\!\!\!\!\!\! C_{i_0\cdots i_{j-1};i_{j+1}\cdots i_{N-1}}
|i_0\cdots i_{j-1} i_{j+1}\cdots i_{N-1}\rangle, 
\end{equation}
so using Eq. (\ref{Pk}) we find that 
\begin{align}
\nonumber
P(\phi;x)&=b_j(x^{2^j}-x_{jm}^{2^j})\sum_{i_l \in\{0,1\},l\not =j }
\!\!\!\!\!\!\! C_{i_0\cdots i_{j-1};i_{j+1}\cdots i_{N-1}}\\
\label{no2toj}
&\times  x^{i_0 2^0+\cdots +i_{j-1} 2^{j-1}+i_{j+1} 2^{j+1}+\cdots +i_{N-1} 2^{N-1}}
\end{align}
where we have assumed that $b_j\not =0$, which is equivalent to $k_j=1$. 
We have also written $(a_j + b_j x^{2^j})=b_j(-x_{jm}^{2^j}+x^{2^j})$. 
Note that $x_{jm}^{2^j}$ is independent of $m$. 
If $b_j=0$, we get an expression which is obtained by multiplying 
the sum of Eq. (\ref{no2toj}) by $a_j$. 
Equation (\ref{no2toj}) shows that the polynomial 
$P(\phi;x)/(x^{2^j}-x_{jm}^{2^j})$ contains only those powers of $x$ which do not have $2^j$ 
in their binary representation and that $x_{jm}$ is a root of $P(\phi;x)$ for each $m=0,1,\ldots ,2^j-1$.   
Therefore, if $k_j=1$, necessary conditions for the qubit $j$ to be unentangled with respect to the rest of the qubits are 
\begin{equation}
\begin{array}{lll}
&\textrm{(IIa) }& P(\phi;x_{jm})=0\textrm{ for every }m=0,1,\ldots ,2^j-1.\\
&\textrm{(IIb) }&    2^j\textrm{ does not appear in the binary representations}\\
&&\textrm{of the exponents of }x \textrm{ in } P(\phi;x)/(x^{2^j}-x_{jm}^{2^j}).
\nonumber
\end{array}
\end{equation}
If $k_j=0$ there is only one condition, namely, 
\begin{equation}
\begin{array}{lll}
&\textrm{(IIc) }&    2^j\textrm{ does not appear in the binary representations}\\
&&\textrm{of the exponents of }x \textrm{ in } P(\phi;x).
\nonumber
\end{array}
\end{equation}
It is easy to see that these are also sufficient conditions. 
The number of unentangled qubits can be obtained by checking 
Conditions (IIa) and (IIb) for every qubit $j$ for which $k_j=1$ 
and Condition (IIc) for the rest of the qubits. 
It is possible to extract information about the number of unentangled qubits 
without using Conditions (IIb) and (IIc), namely, an upper bound for this quantity can be obtained  
by adding to the number of qubits for which (IIa) holds the number of indices $j$ for which 
$k_j=0$. This corresponds to assuming that either (IIb) or (IIc) holds for every qubit.  

\subsection{Example 1}
As an example of the use of this method we consider the state 
given by Eq. (\ref{xi}). Now $k=2^N-1$ and therefore $k_j=1$ for every $j$. 
Equations (\ref{j2}) and (\ref{j0}) together with Condition (IIa) 
show that the number of unentangled qubits is 
at most $N-2$ $(N)$ if $\theta\not =2\pi n$ $(\theta =2\pi n)$. 
In order to simplify the polynomial $P(\xi;x)$ we note that 
\begin{equation}
x^{2^N}-1=(x^2-1)(x^2+1)(x^4+1)\cdots (x^{2^{N-1}}+1).
\end{equation}  
With the help of this and Eq. (\ref{Pxi}) we get 
\begin{align}
\nonumber
&P(\xi^N;x)=(x+1)(x^2+1)(x^4+1)\cdots (x^{2^{N-1}}+1)\\
&+(e^{i\theta}-1)(x^4+1)(x^8+1)\cdots (x^{2^{N-1}}+1)
\end{align} 
Now $x^{2^j}-(x_{jm})^{2^j}=x^{2^j}+1$ when $j\geq 2$ and using the 
above equation one can see that Condition (IIb) holds for $j=2,3,\ldots,N-1$ 
regardless of the value of $\theta$. Furthermore, (IIb) 
holds for every $j$ if $\theta=2\pi n$. In conclusion, 
the qubits $j=2,3,\ldots,N-1$ are always unentangled with respect 
to the rest of the qubits and if $\theta=2\pi n$ the state is separable. 
The same result can be obtained using the reduced single-qubit 
density matrices $\rho_j$. The number of unentangled qubits is equal to 
the number of $\rho_j$ for which $\textrm{det}(\rho_j)=0$. The
values of these determinants have been presented in Example 3 and reproduce 
the aforementioned result.

A necessary step in the calculation of the number of unentangled qubits is to 
apply Condition (IIa) to all qubits $j$ for which $k_j=1$. 
This is equivalent to checking the separability of the state. 
In addition to this, Conditions (IIb) and (IIc) have to be controlled.    
On the other hand, in the case of single-qubit reduced density matrices $\rho_j$,  
the determination of the number of unentangled qubits does not require 
any additional operations in comparison with testing the separability.  
In both cases $\textrm{det}(\rho_j)$ has to be calculated.  
This suggests that the method of reduced single-qubit density matrices is preferable 
if the number of unentangled qubits is studied.

\section{Conclusions}
We have defined a mapping which associates pure $N$-qubit states with a polynomial. 
The roots of this polynomial determine the state completely and vice versa. The structure of the polynomial 
is inspired by the one used in the Majorana representation \cite{Majorana32,Bloch45}. 
The separability of a state can be studied by examining 
the properties of the roots of the corresponding polynomial.  
In particular, we have presented a method which establishes a necessary and sufficient 
condition for a given pure $N$-qubit state $\phi$ to be separable. 
This method provides a new point of view to the pure state separability and 
gives an alternative to the conventional separability test 
of calculating the reduced single-qubit density matrices 
of the state. The separability of $\phi$ can be determined 
by checking whether the numbers $x_{jm}$, defined in equation (\ref{xlm}), 
are roots of the polynomial $P(\phi;x)$ of equation (\ref{poly}). 
Both the numbers $x_{jm}$ and the polynomial $P(\phi;x)$ can be easily obtained  
as a function of the components of the state $\phi$.   
We have illustrated through examples that in some cases the polynomial separability test is 
 easier and faster to apply than the method of reduced single-qubit density matrices.     
We have also shown how the number of unentangled qubits can be obtained 
with the help of the polynomial $P(\phi;x)$. It seems, however, that for this task  
the method of single-qubit density matrices is preferable.

\begin{acknowledgments} 
The authors are grateful to V.I. and M.A. Man'ko for helpful discussions. 
A.M. acknowledges partial support by MIUR Project II04C0E3F3 
Collaborazioni Interuniversitarie ed Internazionali Tipologia C. 
H.M. wants to thank E. Kyoseva, B.W.  Shore, and N.V.  Vitanov 
for comments on an earlier version of the manuscript 
and EC Projects CAMEL and EMALI for financial support. 
\end{acknowledgments}

\end{document}